\documentclass{appolb}
\usepackage{graphicx}

\begin{document}
\title{Magnetic properties in the inhomogeneous chiral phase
\thanks{Presented at Excited QCD 2016}%
}
\author{Ryo Yoshiike, Kazuya Nishiyama, Toshitaka Tatsumi
\address{Department of physics, Kyoto University, Kyoto 606-8502, Japan}
\\
}
\maketitle

\begin{abstract}
We investigate the magnetic properties of quark matter in the inhomogeneous chiral phase, where both scalar and pseudoscalar condensates spatially modulate.
The energy spectrum of the lowest Landau level becomes asymmetric about zero in the external magnetic field,
and gives rise to the remarkably magnetic properties: quark matter has a spontaneous magnetization,
while the magnetic susceptibility does not diverge on the critical point.
\end{abstract}

\section{Introduction}
The phase diagram of QCD has been investigated in the finite chemical potential $(\mu)$ and finite temperature $(T)$ region.
Recently, the existence of the inhomogeneous chiral phase has been newly suggested
and the property of this phase has been actively discussed by the analysis of the chiral effective models \cite{nakano,nickel} or the Schwinger-Dyson approach \cite{mueller} (for a review see Ref.\,\cite{review}).
The observational possibility by the lattice QCD has been also suggested \cite{kashiwa,yoshiike2}.
In this phase, quark condensate has a spatially modulating configuration and such a modulating condensate resembles the FFLO-type superconductivity \cite{ff,lo} or spin/charge density wave \cite{peierls,overhauser}.
Among some configurations of the inhomogeneous quark condensate, we here consider the dual chiral density wave (DCDW) \cite{nakano}, decided by the form, $\Delta({\bf r}) \equiv \langle \bar{\psi} \psi \rangle + i \langle \bar{\psi} i\gamma^5 \tau_3 \psi \rangle = \Delta e^{i{\bf q}\cdot{\bf r}}$, within teh two-flavor QCD.
The DCDW phase is favored compared to other configurations in the 1+1 dimensional system \cite{basar2} or the external magnetic field $(B)$ \cite{nishiyama}.

One may expect that the DCDW phase may be realized in neutron stars
because it is suggested to emerge in the moderate density region by the analysis of the Nambu-Jona-Lasinio (NJL) model \cite{nakano}.
From the observation, the strong magnetic field ($B>10^{12-15}$G) seems to exist in compact stars.
The origin or mechanism of the maintenance of the strong magnetic field have not been fully understood yet
though some mechanisms have been proposed from the macroscopic view point.
It may be important and interesting to investigate the magnetic properties of quark matter to suggest the mechanism from the microscopic theory.


\section{Thermodynamic potential with the weak magnetic field}
We use the two-flavor NJL model with the external magnetic field $(B)$
in the chiral limit. 
Using the DCDW ansatz in the mean field approximation,
the Dirac Hamiltonian is obtained as,
\begin{eqnarray}
 H = -i\vec{\alpha} \cdot {\bf D} - 2G\Delta\gamma^0 \left( \cos qz + i\gamma_5\tau_3\sin qz \right), \label{ham}
\end{eqnarray}
where $\vec{\alpha}\equiv\gamma_0\vec{\gamma}$, ${\bf D} = \nabla + iQ{\bf A}$ and $Q={\rm diag}(e_u,e_d)$ is the electric charge matrix in flavor space.
In the following, $B$ is taken along the $z$ axis, ${\bf A}=(0,Bx,0)$.
According to the Ref.~\cite{frolov}, the energy spectrum is constituted by the Landau levels,
\begin{eqnarray}
 E_{p_zn\zeta\epsilon}^{f} &=& \epsilon\sqrt {\left(\zeta\sqrt{p_z^2+m^2}+q/2 \right)^2 + 2|e_fB|n},~~~(n>0) \\
 E_{p_z\epsilon} &=& \epsilon\sqrt{p_z^2+m^2}+q/2,~~~(n=0) \label{energy}
\end{eqnarray}
where $m\equiv-2G\Delta$ and $\zeta=\pm 1$ denotes the spin polarization.
If $m\geq q/2$, $\epsilon=\pm 1$ corresponds to the positive (negative) energy state, that is, the (anti-)particle state.
However the sign of $\epsilon$ does not always imply that for the lowest Landau level (LLL) ($n=0$) if $m\leq q/2$.
It is worth mentioning that the spectrum of LLL becomes asymmetric about the zero value,
while the one of the higher Landau levels (hLLs) ($n>0$) is always symmetric.

Then the thermodynamic potential takes the form, $\Omega(\mu,T,B;m,q) = \frac{m^2}{4G} + N_c\sum_{f=u,d} \Omega_f$, where,
\begin{eqnarray}
 &\Omega_f = -\frac{|e_fB|T}{4\pi} \int \frac{dp_z}{2\pi} \sum_k \bigg\{ \sum_{n,\zeta,\epsilon} {\rm ln}\left[\omega_k^2 + (E^{f}_{p_zn\zeta\epsilon} - \mu)^2\right] \nonumber \\
 &~~~~~~~~~~~~~~~~~~~~~~~~~~+ \sum_{\epsilon} {\rm ln}\left[\omega_k^2 + (E_{p_z\epsilon} - \mu)^2\right] \bigg\},
\end{eqnarray}
with the Matsubara frequency, $\omega_k=(2k+1)\pi T$.
For the analysis of the response of quark matter in the DCDW phase to the weak external magnetic field,
the thermodynamic potential is expanded about $eB$,
\begin{eqnarray}
 \Omega(\mu,T,B\,;m,q) =& \Omega^{(0)}(\mu,T\,;m,q) + eB\,\Omega^{(1)}(\mu,T\,;m,q) \nonumber \\
                                  &+ (eB)^2\Omega^{(2)}(\mu,T\,;m,q) + \cdots. \label{thermo}
\end{eqnarray}

$\Omega^{(0)}$ corresponds to the thermodynamic potential without $B$ \cite{nakano},
\begin{eqnarray}
 \Omega^{(0)} &=& \frac{m^2}{4G} -2N_c \int \frac{d^3{\bf p}}{(2\pi)^3} \sum_{s=\pm1}  \nonumber \\
  &\times& \left\{ E^{(0)}_s + \frac{1}{\beta}\log\left[ e^{-\beta(E^{(0)}_s-\mu)} + 1 \right]\left[ e^{-\beta(E^{(0)}_s+\mu)} + 1 \right] \right\}, \label{ome0}
\end{eqnarray}
with, $E^{(0)}_s = \left(m^2 + {\bf p}^2 + q^2/4 + sq\sqrt{m^2 + p_z^2}\right)^{1/2}$.
The vacuum part in $\Omega^{(0)}$ should be properly regularized because it includes the UV divergence.

For the analysis of $\Omega^{(1)}$, the chiral anomaly must be considered with caution \cite{yoshiike}. 
According to Refs.~\cite{niemi}, anomalous particle number is generally brought about by spectral asymmetry.
Accordingly anomalous contribution emerges in the thermodynamic potential.
In the DCDW phase, it gives rise to anomalous particle number proportional to $B$ \cite{tatsumi}.
Then $\Omega^{(1)}$ takes the form,
\begin{eqnarray}
\Omega^{(1)}&=&\frac{\mu N_c}{4\pi} \eta_H -\frac{N_c}{4\pi} \int \frac{dp_z}{2\pi} \sum_{\zeta=\pm1} \sum_{\tau=\pm1} \zeta\tau\left( \mu - \tau\omega_\zeta \right) \theta(\tau\omega_\zeta)\theta(\mu - \tau\omega_\zeta) \nonumber \\
&~& -\frac{N_cT}{4\pi} \int \frac{dp_z}{2\pi} \sum_{\zeta=\pm1}\sum_{\tau=\pm1} \zeta\tau\ln\left(1 + e^{-\beta|\omega_\zeta - \tau\mu|} \right), \label{ome1}
\end{eqnarray}
with $\omega_{\zeta} = \sqrt{p_z^2 + m^2} + \zeta q/2$,
which is the odd function about $q$.
The first term represents the contribution of anomaly derived from only LLL
while the second and third term are interpreted as the contributions of valence quarks coming from all the Landau levels.
Note that $\Omega^{(1)}$ does not {\bf diverge} without any regularization.
The anomalous contribution is caused by the spectral asymmetry and the $\eta$-invariant, $\eta_H$, renders,
\begin{eqnarray}
 \eta_H &\equiv& \lim_{s \to +0}\int \frac{dp_z}{2\pi} \sum_{\epsilon}|E_{p_z\epsilon}|^{-s}{\rm sign}(E_{p_z\epsilon}) \nonumber \\
           &=& \left\{ \begin{array}{lc} 
                         -\frac{q}{\pi} & (m>q/2) \\
                         -\frac{q}{\pi} + \frac{2}{\pi}\sqrt{q^2/4 - m^2} & (m<q/2)
                         \end{array} \right..
\end{eqnarray}
Spectral asymmetry can be evaluated with the proper regularization which does not violate the gauge invariance.
Note that in $m>q/2$, this quantity reproduces the Wess-Zumino-Witten (WZW) term effectively derived from the chiral anomaly, argued in Ref.~\cite{son}.
Although the WZW term does not depend on $m$,
the $m$ dependence emerges in our case when $m$ sufficiently becomes small.
Moreover it vanishes in the limit, $m \rightarrow 0$, as should be. 
%

Finally, $\Omega^{(2)}$ does not include the contribution of anomaly and takes the form,
\begin{eqnarray}
 \Omega^{(2)} &=& -\frac{5}{216\pi} \int \frac{dp_z}{2\pi} \sum_{\zeta=\pm1} \frac{1}{\omega_\zeta} \left[ \frac{1}{e^{\beta(\omega_\zeta + \mu)}+1} - \frac{1}{e^{\beta(\omega_\zeta - \mu)}+1} - 1\right].
\end{eqnarray}
There is still included the UV divergence.

The values of the order parameters, $m$ and $q$, are determined for each $\mu$, $T$ and $B$ and make the thermodynamic potential minimum.
They can be expanded about $eB$,
\begin{eqnarray}
 m(\mu,T,B) &=& m^{(0)}(\mu,T) + eB m^{(1)}(\mu,T) + (eB)^2 m^{(2)}(\mu,T), \\
 q(\mu,T,B) &=& q^{(0)}(\mu,T) + eB q^{(1)}(\mu,T) + (eB)^2 q^{(2)}(\mu,T).
\end{eqnarray} 
The minimized thermodynamic potential is represented as,
$\Omega^{\rm min}(\mu,T,B) \equiv \Omega(\mu,T,B;m,q)|_{m=m(\mu,T,B),q=q(\mu,T,B)}$.

\section{Magnetic properties}

\subsection{Spontaneous magnetization}
Magnetization is defined as the first derivative of the thermodynamic potential about the magnetic field.
From the extremum conditions, $\partial \Omega / \partial m,q = 0$, the spontaneous magnetization ($M_0$) takes the form,
\begin{eqnarray}
 M_0(\mu,T) \equiv -\frac{\partial \Omega^{\rm min}}{\partial B}\bigg|_{B\rightarrow0} = -e\Omega^{(1)}(\mu,T;m=m^{(0)}, q=q^{(0)}).
\end{eqnarray}
From the Eq.\,(\ref{ome1}), $\Omega^{(1)}$ has a finite value only when $m^{(0)}$ or $q^{(0)}$ does not vanish.
In other words, $M_0$ does not vanish only in the DCDW phase.
Furthremore, it includes the contribution of not only anomaly but also valence quarks.

\begin{figure*}[ht]
 \centering
  \begin{tabular}{c}
      \begin{minipage}{0.5\hsize}
        \begin{center}
          \includegraphics[width=6cm]{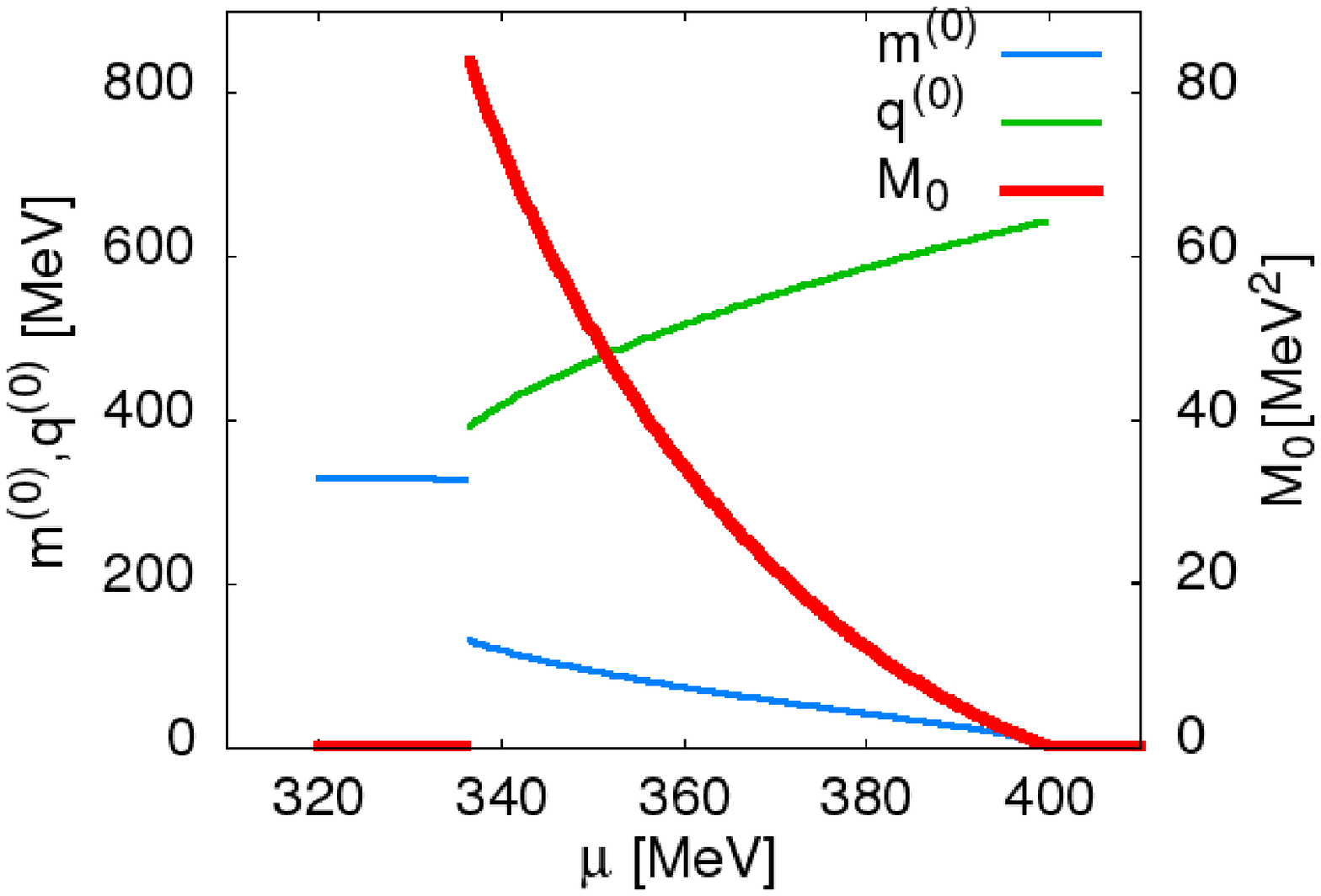}
          (a) $T=0$
        \end{center}
      \end{minipage}

      \begin{minipage}{0.5\hsize}
        \begin{center}
          \includegraphics[width=6cm]{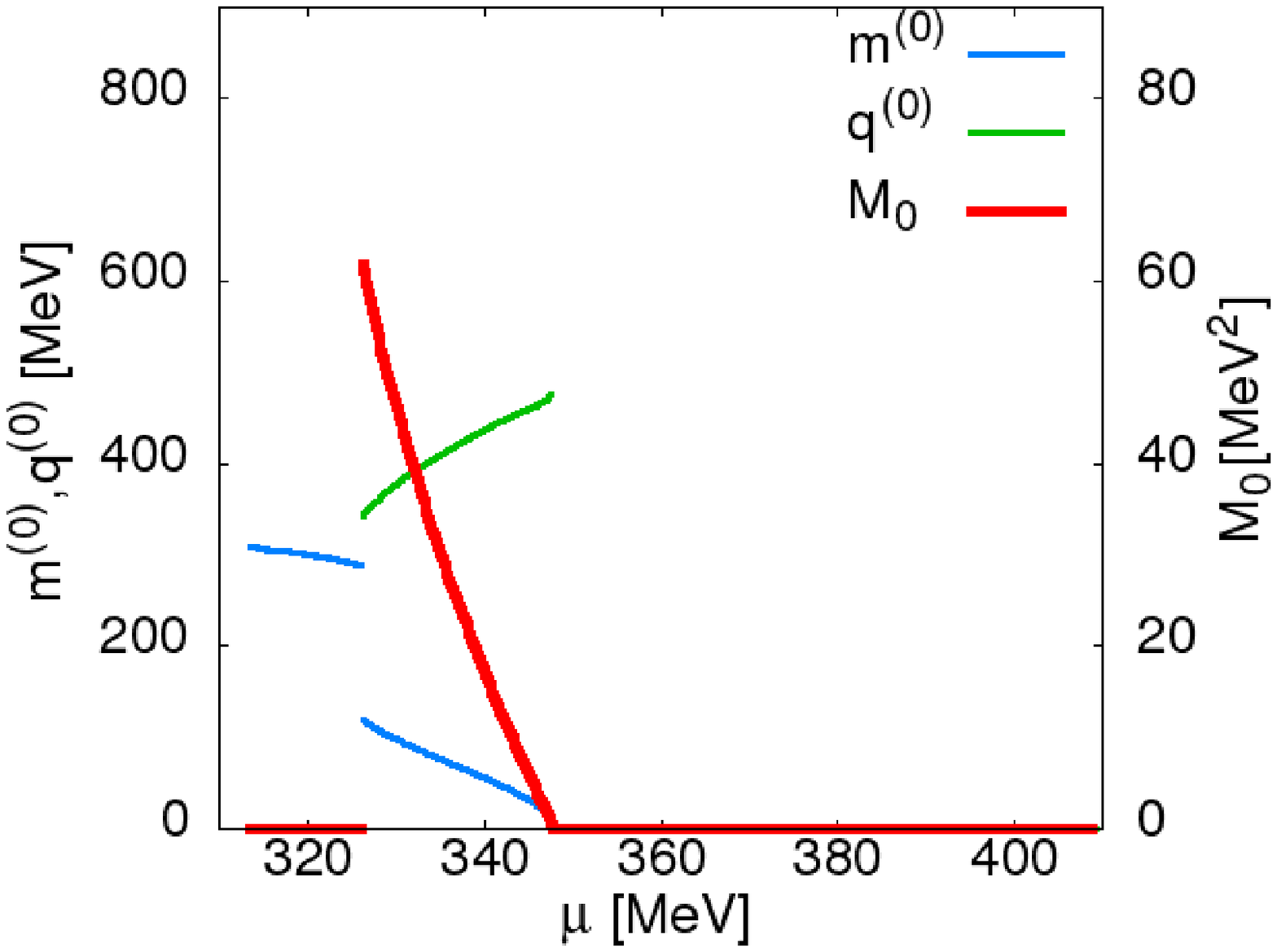}
          (b) $T=30{\rm MeV}$
        \end{center}
      \end{minipage}      
 \end{tabular}
 \caption{The chemical potential dependence of the order parameters and spontaneous magnetization. The DCDW phase exists in the $\mu$ region bounded by the first-order and the second-order phase transition points.}
 \label{spo}
\end{figure*}

The Fig.\,\ref{spo} shows the $\mu$ dependence of $M_0$ at $T=0, 30$MeV.
As temperature increases, the range of $\mu$ gets narrow and the magnitude of $M_0$ decreases.
Assuming a sphere of quark matter with constant density ($\mu=340$MeV) and uniform magnetization, the magnitude of the magnetic field made from $M_0$ is estimated, $B_{\rm mag} = \frac{8\pi}{3}M_0 \sim 10^{16}$G, on the surface.

\subsection{Magnetic susceptibility}

\begin{figure}[hb]
 \centering
  \includegraphics[width=6cm]{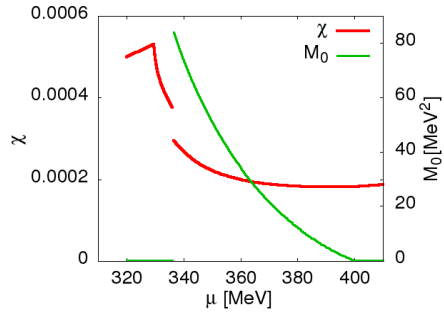}
 \caption{The chemical potential dependence of normalized magnetic susceptibility and spontaneous magnetization at zero temperature.}
 \label{sus}
\end{figure}

Magnetic susceptibility ($\chi$) is defined as the first derivative of the magnetization about the magnetic field.
From the stationary conditions, $\chi$ takes the form,
\begin{eqnarray}
  \chi(\mu,T) &\equiv& - \frac{\partial^2 \Omega^{\rm min}}{\partial B^2}\bigg|_{B\rightarrow0} \nonumber \\
  &=& - e^2\Big( 2\Omega^{(2)}(\mu,T;m,q) + m^{(1)}\partial_m \Omega^{(1)}(\mu,T;m,q) \nonumber \\
  &~&~~~~~~~~~~~~~~~~~~~~+ q^{(1)} \partial_q \Omega^{(1)}(\mu,T;m,q) \Big)\Big|_{m=m^{(0)},q=q^{(0)}},
\end{eqnarray}
which has the UV divergence coming from $\Omega^{(2)}$.
Therefore it should be normalized to satisfy the condition, $\chi(\mu=0,T=0)=0$.
Thus the normalized $\chi$ is defined by subtracting the vacuum one, $\chi^{\rm nom}(\mu,T) \equiv \chi(\mu,T) - \chi(\mu=0,T=0)$.
Fig.\,\ref{sus} shows the $\mu$ dependence of $\chi^{\rm nom}$, which exhibits some singular behavior.
The cusp behavior reflects the singularity of the thermodynamic potential on the point where valence quarks begin to appear.
Quark matter undergoes the first order phase transition at the critical point, where order parameters are discontinuous (Fig.\,\ref{spo}) and $\chi^{\rm nom}$ is also discontinuous. 
However, $\chi^{\rm nom}$ does not indicate any singularity on the second order critical point (Fig.\,\ref{spo})
while, in the Ising model, magnetic susceptibility diverges on the second order critical point.
Therefore such behavior is unique in the case of the DCDW phase transition.

\section{Summary}
We have investigated the magnetic properties of quark matter in the inhomogeneous chiral phase.
Spontaneous magnetization, which has both the contributions of anomaly and valence quarks, emerges there.
The magnitude of magnetic field made from it may be estimated to be comparable with that in the observation of magnetars.
Magnetic susceptibility does not diverge on the second order phase transition point.
The behavior is different from one in the familiar ferromagnetic system and unique in the inhomogeneous chiral phase transition.
One of the author (RY) thanks C. Providencia for her hospitality during the stay in Coimbra.





\end{document}